# Insights into bubble droplet interactions in evaporating polymeric droplets

Gannena K S Raghuram[1#], Durbar Roy[1#], D Chaitanya Kumar Rao[2], Aloke Kumar[1*], Saptarshi Basu[1**]

[1]Department of Mechanical Engineering, Indian Institute of Science, India

[2]Department of Aerospace Engineering, Indian Institute of Technology Kanpur, India

*Corresponding author email: alokekumar@iisc.ac.in

**Corresponding author email: sbasu@iisc.ac.in

# Equal contributions

## Abstract

Polymer droplets subjected to a heated environment have significance in several fields ranging from spray drying and powder formation to surface coating. In the present work, we investigate the evaporation of a high viscoelastic modulus aqueous polymeric droplet in an acoustically levitated environment. Depending on the laser irradiation intensity, we observe nucleation of a bubble in the dilute regime of polymer concentration, contrary to the previously observed bubble nucleation in a semi-dilute entangled regime for low viscoelastic modulus polymer droplets. After the bubble nucleation, a quasi steady bubble growth occurs depending on the laser irradiation intensity and concentrations. Our scaling analysis reveals that bubble growth follows Plesset-Zwick criteria independent of the viscoelastic properties of the polymer solution. Further, we establish that the onset of bubble growth has an inverse nonlinear dependence on the laser irradiation intensity. At high concentrations and laser irradiation intensities, we report the expansion and collapse of polymer membrane without rupture, indicating the formation of an interfacial skin with significant strength. The droplet oscillations are primarily driven by the presence of multiple bubbles and, to some extent, by the rotational motion of the droplet. Finally, depending on the nature of bubble growth, different types of precipitate form contrary to the different modes of atomization observed in low viscoelastic modulus polymer droplets.

## 1.Introduction

Polymer droplet and thin polymer film evaporation continue to evoke scientific curiosity in various applications ranging from targeted drug delivery, thin films, and coatings to surface patterning (Wilms 2005; Pathak and Basu 2016a). Understanding the kinetics and dynamics of evaporating polymer droplets is crucial for practical applications. Polymer droplet evaporation involves complex events such as solvent evaporation, subsequent build-up of polymer concentration at the air-liquid interface, and precipitate formation (Littringer et al. 2012; Al Zaitone et al. 2020). Depending on the initial polymer concentration and solvent evaporation rate, the accumulation of solute (polymer) at the surface aids in forming a gel-type layer, also called the skin layer (Okuzono et al. 2006; Pauchard and Allain 2003b; Pauchard and Allain 2003c). Based on the properties of the polymer and drying kinetics of droplets, the final morphology of polymer residue can be in the form of a wrinkled pattern (Pauchard and Allain 2003b), buckled structure (Pauchard and Allain 2003c), smooth solid precipitate (Raghuram et al. 2021), or ring pattern (Raghuram et al. 2021). Investigations on the evaporating polymer droplet have been performed in a contact environment (hydrophilic substrates) under natural drying conditions (Baldwin and Fairhurst 2014; Mamalis et al. 2015; Baldwin et al. 2011; Baldwin et al. 2012; Pauchard and Allain 2003a). Pauchard et al.(Pauchard and Allain 2003a) revealed the skin layer formation near the vapor/drop interface in evaporating sessile droplets. It was demonstrated that as the enclosed liquid volume decreases, the skin layer deforms, leading to buckling instability in the droplet. Depending on the experimental conditions, different shape instabilities have been reported, from buckled structure to the wrinkled pattern on the droplet surface. Baldwin et al. (Baldwin and Fairhurst 2014; Baldwin et al. 2011; Baldwin et al. 2012) and Mamalis et al.(Mamalis et al. 2015) explored the influence of molecular weight and concentration on final deposit formation. Depending on the competing effect between advective polymer build-up and diffusive flux near the three-phase contact line, pillars and puddle-like deposits on glass surfaces have been observed.

The bubble dynamics in multi-component droplets through external heating are extensively reported across various experimental configurations (Mura et al. 2014; Rao et al. 2018; Pathak and Basu 2016b; Rao et al. 2017; Miglani et al. 2014; Antonov and Strizhak 2019; Restrepo-Cano et al. 2022). In particular, Rao et al.(Rao et al. 2018; Rao et al. 2017) studied bubble dynamics and breakup mechanisms dynamics in burning multi-component miscible droplets. Different modes of bubble-induced droplet shape oscillations were reported depending on the size of the bubble, volatility differential, and concentration of the components. It was shown that a significantly larger volatility difference leads to severe shape oscillations induced by the breakup of a large bubble, whereas lower volatility differential results in mild shape oscillations caused by the breakup of a relatively small bubble.

Most experimental studies have been performed primarily to understand the dynamics of evaporating polymer droplets in a contact environment (hydrophilic substrates), and the literature on the evaporation of isolated polymer droplets (non-contact environment) of broad viscoelastic natures is scarce. The contact-free environment is provided by a relatively simplistic methodology of acoustic levitation (Rao et al. 2022; Gannena et al. 2022), which allows one to accurately capture the droplet shape oscillations and short spatio-temporal instabilities at the vapor-liquid interface (Gonzalez Avila and Ohl 2016).

In the context of acoustic levitation, Rao et al. (Rao and Basu 2020; Rao et al. 2020) investigated the dynamics of levitated emulsion droplets under external radiative heating. They reported that droplet breakup is categorized into three types: breakup through bubble growth, sheet breakup, and catastrophic breakup, depending on the onset of vapor bubble nucleation. It is also shown that the size of secondary droplets depends on the mode of droplet breakup. In the case of nanoparticle-laden droplets, Pathak et al. (Pathak and Basu 2016b) studied how nanoparticles could affect the dynamics of fuel droplets under external radiative heating. During evaporation, the accumulation of nanoparticles through orthokinetic aggregation leads to the formation of nanoparticle aggregates. These aggregates act as nucleation sites leading to heterogeneous boiling inside the droplet and subsequent breakup of parent droplets.

Previously, we investigated the coupled effect of the skin layer and bubble in evaporating low viscoelastic modulus polymer (Polyacrylamide) droplets under a heated environment. During evaporation, bubble nucleation in the semi-dilute entangled regime of polymer concentration results in membrane growth, followed by its rupture at low to medium irradiation intensities. At high irradiation intensities, the PAM droplets undergo universally observed ligament-mediated and sheet breakup (Gannena et al. 2022). However, it is essential to understand how the bulk viscoelasticity of polymer droplets can affect the underlying bubble and droplet dynamics. The current study explores the nature of skin layer and bubble interaction on the droplet dynamics in evaporating high viscoelastic modulus polymer (PEO) droplets. Note that the motivation of the present theoretical framework is to provide the approximate scales for various physical quantities observed during the experiments and explore the physics of the phenomenon. The exact analytical or numerical solutions of the coupled governing equations of momentum, heat and mass transfer is outside the scope of the present study.

This paper is organized as follows. Section 2 details materials and methods involving polymer solution preparation and its properties (§ 2.1), and experimental methodology (§ 2.2). The results and discussion involve global observations (§ 3.1), evaporation (§ 3.2), steady bubble growth, membrane dynamics (§ 3.3), droplet shape oscillations and precipitate formation (§ 3.4). The conclusions of the present study are provided in § 4.

## 2. Materials and methods

### 2.1. Polymer solution preparation and properties

Various concentrations of PEO (Sigma-Aldrich) solutions ranging from 0.06 to 2% (w/w) of molecular weight $M_W$ of $4 \times 10^6$ g mol$^{-1}$ are prepared by dissolving Polyethylene oxide (PEO) powder in DI water. PEO solutions are stirred at 600 rpm for 24 hours to ensure proper mixing. The preparation methodology of PAM solutions can be found in (Gannena et al. 2022). Throughout the current experimental study concentration of the polymer solution ($c$) is normalized with overlap concentration ($c^*$). The value of the critical overlap concentration of PEO is $c^* = 0.071\%$ ($w/w$). It is obtained by applying the Flory relation $c^* = \frac{1}{[n]}$, where $[n] = 0.072 M_W^{0.65}$ is obtained from the Mark-Houwink-Sakurada correlation (Tirtaatmadja et al. 2006). Entanglement concentration for the current polymer obtained as $c_e = 0.42\%$ w/w by using the relation $c_e = 6c^*$. The solutions having concentration ratios $c/c^* < 1, 1 < c/c^* < c_e/c^*$ and $c/c^* > c_e/c^*$ are dilute regime, semi-dilute un-entangled regime, and semi-dilute entangled regime, respectively. More information on the definition of regimes of polymer concentrations and non-dimensionalization of concentrations for PAM can be found in (Gannena et al. 2022). To confirm the viscoelastic nature of PEO and PAM solutions, rheological tests are performed on a rheometer (Anton Paar, MCR702) with cone and plate geometry. The diameter and angle of the cone and plate are 50 mm and 1º, respectively. Figure 1 shows the inherent viscoelastic nature of PAM and PEO bulk solutions. Here $G'$ and $G''$ represent storage and loss modulus, respectively. It can be seen that the storage and loss modulus of the PAM solution are in the $O\ (10^{-2} - 10^{-1})\ Pa$, whereas for PEO storage and loss modulus are in the $O\ (10^1)\ Pa$. The viscoelastic modulus of the polymer solutions depends on the entanglements present in the solution. A higher value of entanglements corresponds to a higher viscoelastic modulus of the polymer solution. The quantitative criteria governing the entanglements in each polymer solution is given by entanglement density $N_e$. It is given as

$$N_e = (M_W/M_e)(c/c^*) \qquad (2.1)$$

Here $M_W$ and $M_e$ represents the molecular weight and entanglement molecular weight of the polymer, respectively. The $M_e$ value for PEO and PAM is 2000 g mol⁻¹ (Rubinstein and Colby 2003) and 9000-23000 g mol⁻¹ (Plastics Technolgy) respectively. Although $M_W$ and $c/c^*$ are in same range for PAM and PEO, the lower value of $M_e$ for PEO gives higher value of $N_e$ for PEO compared to PAM. However, the current study focuses on exploring the effect of laser heating in high viscoelastic modulus polymer droplets without dwelling too much on the rheological differences between PAM and PEO solutions.

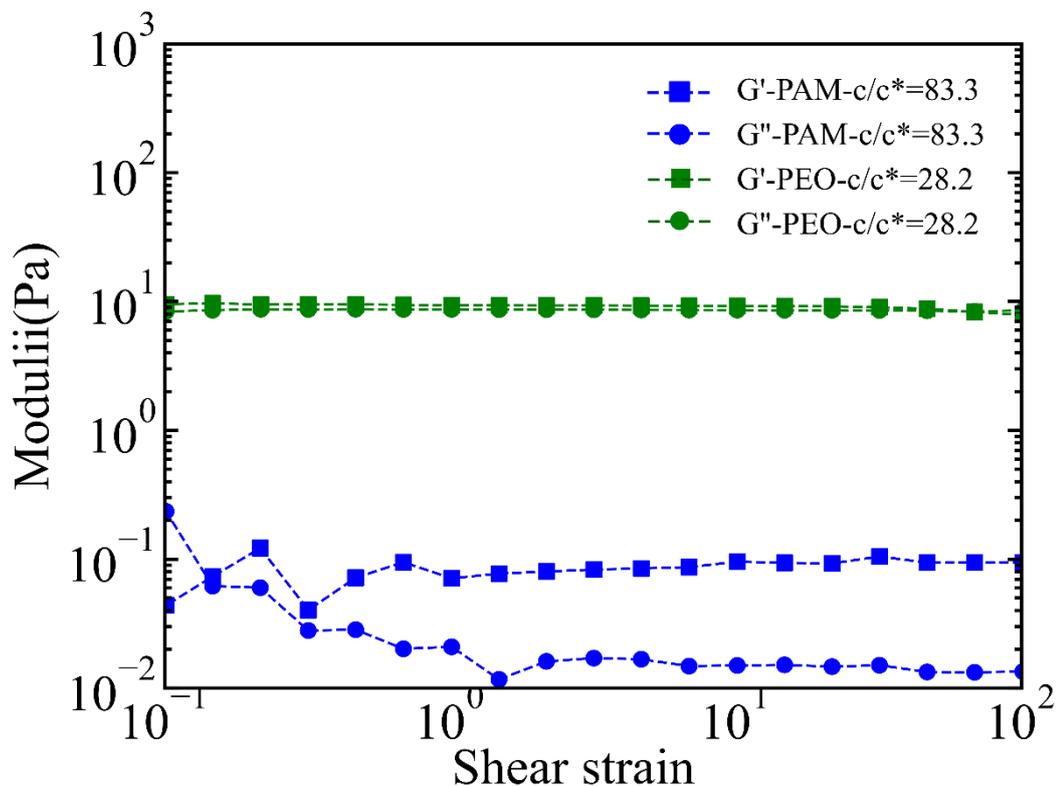

**Figure 1**. Behaviour of storage and loss modulus with shear strain. Here $G'$ and $G''$ represent storage and loss modulus, respectively.

## 2.2. Experimental methodology

Figure 2 depicts the experimental setup used in the current study. The droplets of PEO solutions comprising different concentrations are levitated using a single-axis acoustic levitator (Tec5) with 100 kHz frequency. The droplets are externally heated with a tunable continuous $CO_2$ laser (Synrad 48, wavelength ∼10.6 µm, max power ($P_{max}$) ∼ 10 W) with a beam diameter of 3.5 mm. A high-speed camera (Photron SA5) and a high-speed laser for illumination (CAVILUX® Smart UHS, 640 nm) are used to capture the droplet evaporation and oscillation processes. The high-speed images are recorded at 10000 fps, and the spatial resolution of the recorded images is 6.7 µm/pixel. The recorded grey scale images were contrast enhanced and converted to binary images. The droplet shape was then reconstructed using an edge detection methodology to obtain its maximum horizontal and vertical lengths $D_H$ and $D_V$, respectively. The equivalent diameter of the droplet is calculated using the relation, $D = \sqrt[3]{D_H^2 D_V}$ . The above-mentioned measurement is performed using "Analyze particle" plugin in the "ImageJ (version 2.0)" software. The approximate diameter of the droplets used in the current study is 0.95 ± 0.05 mm. After evaporation, the precipitates are examined using a scanning electron microscope (SEM) (VEGA3, TESCAN) at EHT of 5KV, using a secondary electron detector. In the current study, the irradiation intensity from the laser is non-dimensionalized with the enthalpy of vaporization. More details on non-dimensionalization can be found in Gannena et al.(Gannena et al. 2022). An Infrared camera is used to measure the evolving droplet surface temperature with time. The IR camera (FLIR SC5200: pre-calibrated for a standard emissivity of 1 with an accuracy of ±1 °C) is operated at 50 frames per second (fps) with a spatial resolution of 6.42 µm/pixel. It has been reported that the emissivity for water is between 0.95 and 0.98 (Mikaél'A 2013; Wolfe and Zissis 1978). The change in temperature due to the change in emissivity is 0.03 °C, which is assumed to be negligible. The captured images are processed using ALTAIR software (FLIR Systems, version 5.91.10.797) to extract the droplet temperature during the heating process. The temperature information is gained by defining a linear region of interest along the droplet diameter in each IR frame, and the maximum temperature on the surface of the droplet is calculated. The temperature at the droplet's core is anticipated to be higher than at the surface due to the volumetric nature of the absorption process (Abramzon and Sazhin 2006). Further, the temperature at the interface is lower due to evaporative cooling effects.

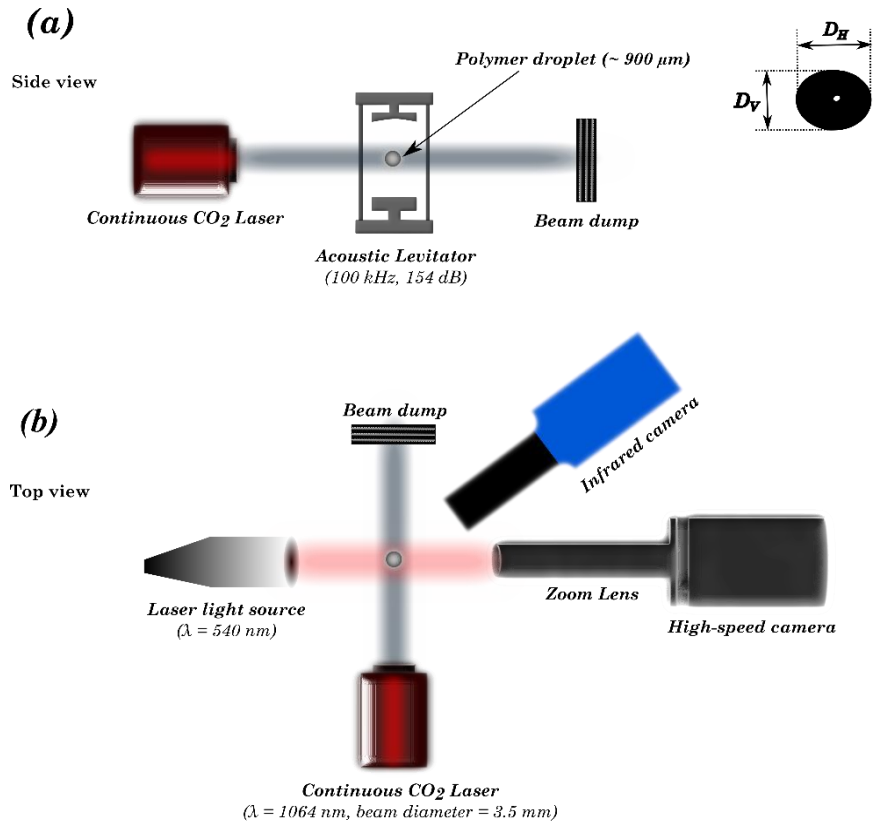

**Figure 2**. Schematic representing (a) Side view and (b) top view of the experimental setup. The droplet is levitated using a single-axis acoustic levitator and evaporated using a continuous $CO_2$ laser. The droplet evaporation and bubble dynamics are captured with a high-speed camera, and a pulsed laser light source provides the backlighting. The surface temperature of the droplet is captured using an IR camera. $D_H$ and $D_V$ are the maximum horizontal and vertical lengths of droplets, respectively.

## 3. Results and discussions

### 3.1. Global observations

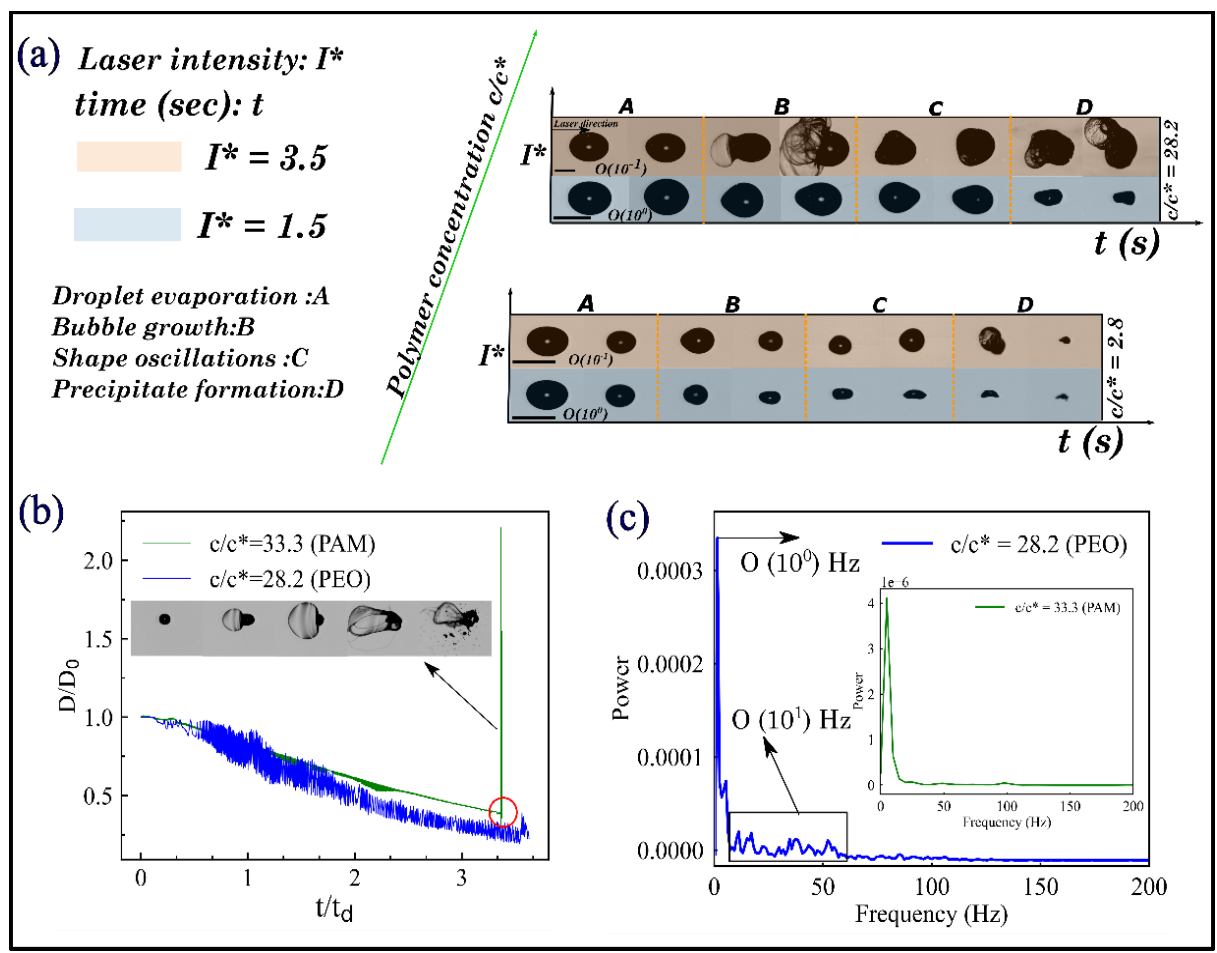

**Figure 3**. (a) Global observations of the droplet evaporation and bubble dynamics associated with PEO aqueous solutions. The influence of polymer concentration and laser irradiation intensity is shown. Here, O () symbol indicates the order of time scale of occurrence. The scale bar represents 1 mm. (b) Temporal variation of normalized diameter of polymer droplet at $I^* = 1.5$ (c) Power spectrum density of diameter of polymer droplet at $I^* = 1.5$.

A global overview summarizing the interaction between a continuous laser and an acoustically levitated droplet for a range of polymer concentrations and irradiation intensity is shown in figure 3. For concentrations above the entangled regime ($c/c^* = 28.2$) and at a high irradiation intensity $I^* = 3.5$, after a period of smooth evaporation (Phase A), we observe bubble growth (Phase B) followed by shape oscillations and precipitate formation (Phases C and D). The bubble growth observed in Phase B can be attributed to asymmetric membrane growth (Gannena et al. 2022), which occurs in the early stages of evaporation with growth scales of $O\ (10^{-4})\ s$. Phases C and D are dominant in the later stages of evaporation. Similar phases are observed at $c/c^* = 28.2$ for $I^* = 1.5$ and $c/c^* = 2.8$ for all the irradiation intensities. The observations indicate that the dynamics in phases B, C and D are significantly affected by bubble growth depending on irradiation intensity and polymer concentration. The observed bubble dynamics differ from the previously reported membrane development, rupture, and breakup in evaporating low viscoelastic modulus (PAM) droplets (Gannena et al. 2022). Figure 3(b) shows the variation of normalized diameter ($D/D_0$) with normalized time ($t/t_d$) at $I^* = 1.5$ for PAM and PEO. Here, $t_d$ denotes the time scale for thermal diffusion and is defined as

$$t_d = \frac{R_0^2}{\alpha_l} = \frac{\rho_l c_p R_0^2}{k} \qquad (3.1)$$

Where $R_0$ is the initial radius of the droplet, $\alpha_l$ is the thermal diffusion coefficient, $\rho_l$ is the density of liquid, $c_p$ is the specific heat capacity, and $k$ is the thermal conductivity of the liquid. The interaction of a typical PAM droplet with an IR laser in an acoustically levitated field consists of smooth evaporation, nucleation of bubble, bubble expansion, rupture of viscoelastic membrane and subsequent fragmentation of the polymeric droplet through various pathways (see inset figure 3(b)). The dynamics is significantly different from the PEO droplets. The regression data for $c/c^* = 28.2$ and $I^* = 1.5$ (PEO droplet) encapsulates phenomena like droplet evaporation, bubble growth, and droplet rotation, which is visible due to the entrapped bubble. A power spectrum is obtained for the evaporating polymer droplet to understand whether we can decompose the evaporation curve into its constituent physical components (see figure 3(c)). For evaporating PEO droplets at $c/c^* = 28.2$ and $I^* = 1.5$, we observe two frequency bands of $O(10^0)$ Hz and $O(10^1)$ Hz. These frequency bands indicate the evaporation and rotational frequencies of the droplet, respectively. Note that the high-frequency band of $O(10^1)$ Hz is absent for evaporating PAM droplets, confirming that rotational dynamics are absent (see inset figure 3(c)). Furthermore, bubble nucleation occurs for $c/c^* > 10$ in a semi-dilute entangled regime for evaporating PAM droplets (Gannena et al. 2022). However, bubble nucleation occurs even in the dilute regime for evaporating PEO droplets. The essential physical mechanisms and their theoretical scales will be elucidated in subsequent sections.

## 3.2. Droplet evaporation (Phase A)

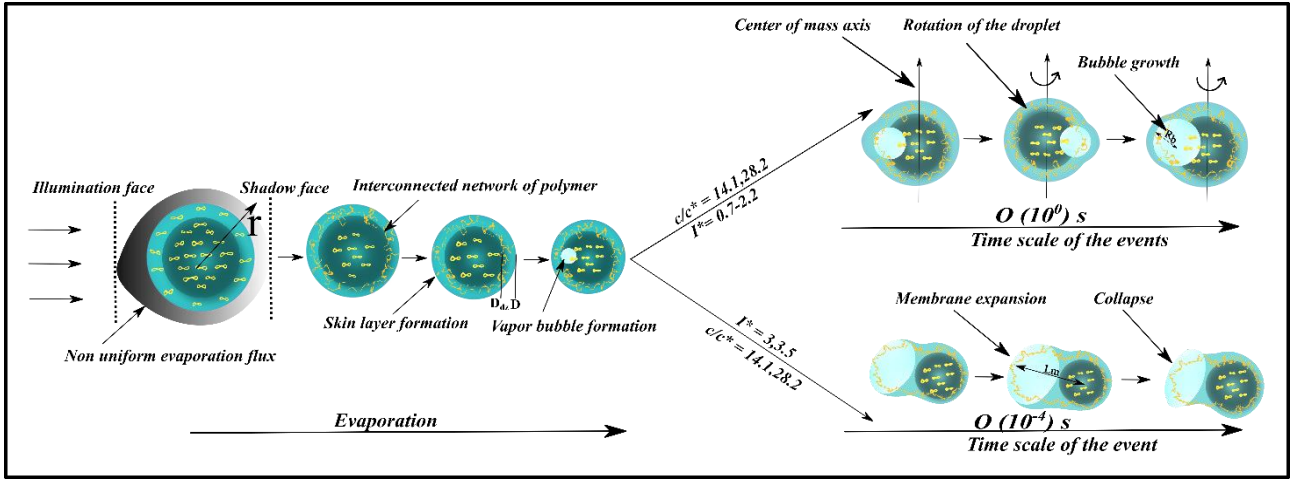

**Figure 4** Illustration of skin layer formation and bubble dynamics in a typical evaporating polymeric droplet

The formation and subsequent dynamics of the vapor bubble during the evaporation of the high viscoelastic modulus polymeric droplet PEO is shown schematically in figure 4. As the polymeric droplet evaporates, the accumulated polymer along the interface forms an interconnected network of polymer, which undergoes a phase transition into a skin layer at a critical gelation polymeric concentration. This is followed by subsequent bubble nucleation as evaporation proceeds (Gannena et al. 2022). At low irradiation intensities, we observe the growth of a stationary bubble while protruding the droplet (see figure 4 and supplementary Movie2). At higher irradiation intensities, membrane expansion and collapse are observed (see figure 4 and supplementary Movie1 ). The membrane expansion and collapse without rupture signifies the creation of a significantly higher-strength skin layer compared to previously observed membrane growth, rupture and breakup in evaporating polyacrylamide (PAM) droplets (Gannena et al. 2022). The quantitative criteria for the formation of the skin layer is given by Peclet number ($Pe$). The Peclet number is defined as $Pe = t_{dp}/t$, where $t_{dp}$ represents the diffusion time scale of polymer molecules inside the droplet and $t$ indicates the evaporation time scale of the droplet. Here, $t_{dp} = D_0^2/D_P$, where $D_0$ represents the initial diameter of the polymer droplet and $D_P$ represents the self-diffusion coefficient of the polymer molecule. The self-diffusion coefficient of the polymer is defined as

$$D_P = \frac{K_B T_R}{6\pi\mu\varepsilon} \qquad (3.2)$$

where correlation length is defined as

$$\varepsilon = R_g \left(\frac{c}{c^*}\right)^{\frac{\vartheta}{1-3\vartheta}} \qquad (3.3)$$

Here $K_B$, $T_R$, and $\mu$ denote the Boltzmann constant, room temperature, and dynamic viscosity of the solvent, respectively. Excluded volume coefficient ($\vartheta$) is set to 0.588 (Raghuram et al. 2021). In the present experimental investigation, $Pe \gg 1$ for all the irradiation intensities and it has been previously demonstrated that $Pe \gg 1$ results in the creation of a skin layer (Gannena et al. 2022). The skin layer thickness $h_0$ is calculated using the conservation of mass of polymer in the liquid droplet. It is defined as

$$h_0 = 0.5 \left( D - \left( D_0 \left( \frac{\widetilde{D}^3 \widetilde{\rho} \emptyset_g - \emptyset_p}{\widetilde{\rho} \emptyset_g - \emptyset_p} \right)^{1/3} \right) \right) \tag{3.4}$$

where $\emptyset_p$ and $\emptyset_g$ represent initial polymer mass fraction and gelation mass fraction, respectively. $\widetilde{\rho}$ indicates the ratio of polymer density to liquid density. $\emptyset_g$ is assumed to be approximately equal to 1, where the polymer concentration is the highest throughout the droplet. More details on equation (3.4) can be referred from our previous work(Gannena et al. 2022). The skin layer thickness ($h_0$) is used for explaining the membrane dynamics (see § 3.3). Depending on the concentration regime, the bubble nucleation differs significantly for low viscoelastic modulus (PAM) and high viscoelastic modulus (PEO) droplets. Figure 5 compares high-speed images of evaporating PEO and PAM droplets for $c/c^* = 2.8$ and $c/c^* = 3.3$ (at $I^* = 2.2$), respectively. Here, both the chosen concentrations fall inside the semi-dilute unentangled regime of polymer solutions. At $c/c^* = 3.3$, for a fluid with a low viscoelastic modulus (PAM), the droplet evaporates without bubble nucleation (figure 5(b)). However, for high viscoelastic modulus (PEO) droplets, after a period of evaporation, a bubble starts to grow at around $t = 1500\ ms$ (figure 5(a)). This can be further corroborated by the temporal variation of non-dimensional diameter ($D/D_O$) with non-dimensional time ($t/t_d$) (see figure 5(c)). The pronounced oscillatory behaviour in the evaporation curve for PEO at $c/c^* = 2.8$ and $I^* = 2.2$ indicates droplet rotation. However, the evaporation curve is smooth for PAM at $c/c^* = 3.3$. The entanglement density $N_e$ increases more steeply for PEO than PAM (see figure 5(d)). PEO leads to more entanglements than PAM, even at semi-dilute and dilute limits. This perhaps leads to the skin layer formation at a much earlier concentration regime (dilute and semi-dilute unentangled regime) and subsequent bubble nucleation at a given irradiation intensity. See figure S3 for further evidence of bubble nucleation in dilute and semi-dilute unentangled regimes of polymer concentrations in PEO fluid droplets at a specific irradiation intensity.

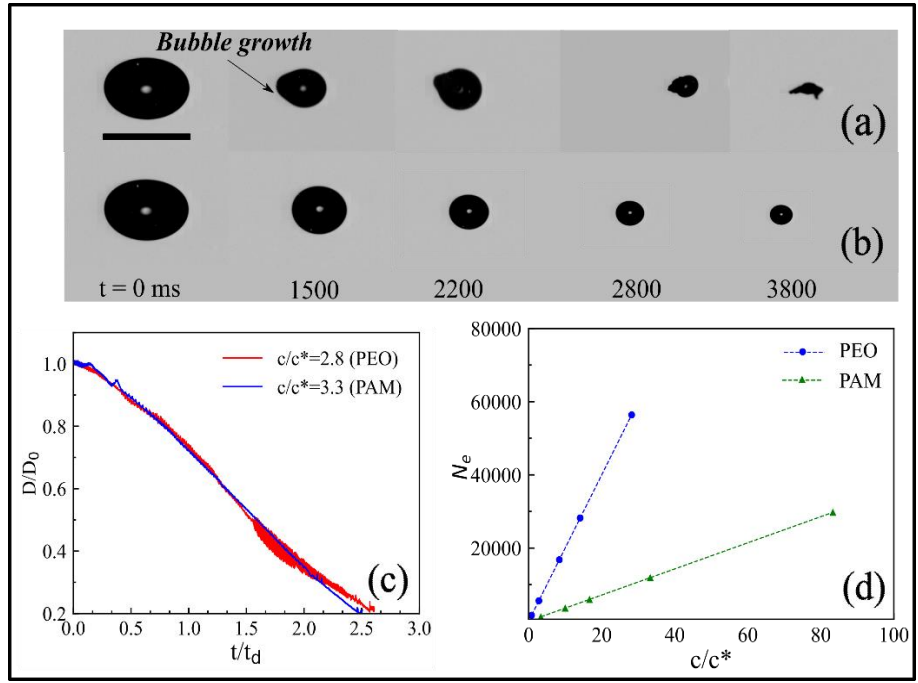

**Figure 5** (a) High-speed snapshots of PEO droplet at $c/c^* = 2.8$ and $I^* = 2.2$ (b) High-speed snapshots of PAM droplet at $c/c^* = 3.3$ and $I^* = 2.2$. The scale bar indicates 1 mm. (c) Temporal variation of drop diameter at $I^* = 2.2$ (d) Variation of entanglement density with non-dimensional concentration.

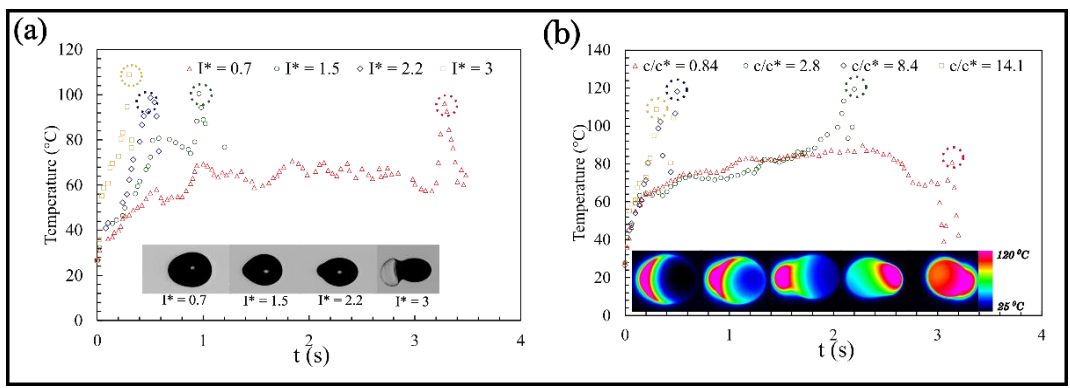

**Figure 6** Temporal evolution of the droplet surface temperature corresponding to (a) $c/c^* = 14.1$ (b) $I^* = 3$. The dotted circle represents the bubble expansion after bubble nucleation. In figure (a), the inset represents high-speed snapshots of bubble expansion after bubble nucleation. The inset in figure (b) represents experimental IR images of evaporating polymer droplets at $c/c^* = 14.1$ and $I^* = 3$.

The experimental temporal surface temperature evolution of a polymer droplet and the experimental evidence of bubble expansion on the illuminating face of the droplet (towards laser direction) is provided through IR thermography. Figure 6 displays the time evolution of the surface temperature of evaporating PEO droplets under varying irradiation intensities and concentrations. At a concentration, $c/c^* = 14.1$ and irradiation intensities $I^*$ ranging from 0.7−3 (see figure 6(a)), surface temperature of the polymer droplet rises with time, reaches a saturation limit, and then abruptly peaks. The peak temperature corroborates the bubble expansion close to the skin layer of the evaporating polymer droplet (see inset figure 6(a)). At higher irradiation intensities, the surface temperature of the polymer droplet rapidly increases, and vice versa at lower irradiation intensities. Also, the peak temperature is attained much earlier for higher irradiation intensities compared to lower irradiation intensities indicating that the bubble nucleation occurs significantly earlier in the droplet's lifetime (assuming negligible time difference between nucleation and discernible bubble).

At an irradiation intensity of $I^* = 3$ irrespective of the concentrations ($c/c^*$) ranging from 0.84-14.1), the initial increase in surface temperature of the polymer droplet remains nearly constant. However, the peak temperature is attained much quicker for higher concentrations compared to lower concentrations, indicating that bubble nucleation occurs much early (see figure 6(b)). Further, evidence of bubble nucleation close to the illuminating face of the droplet can be explained through experimental IR images (see inset figure 6(b)). We can observe a high-temperature zone (>$100^0$ C), further confirming bubble nucleation and expansion on the illuminating face.

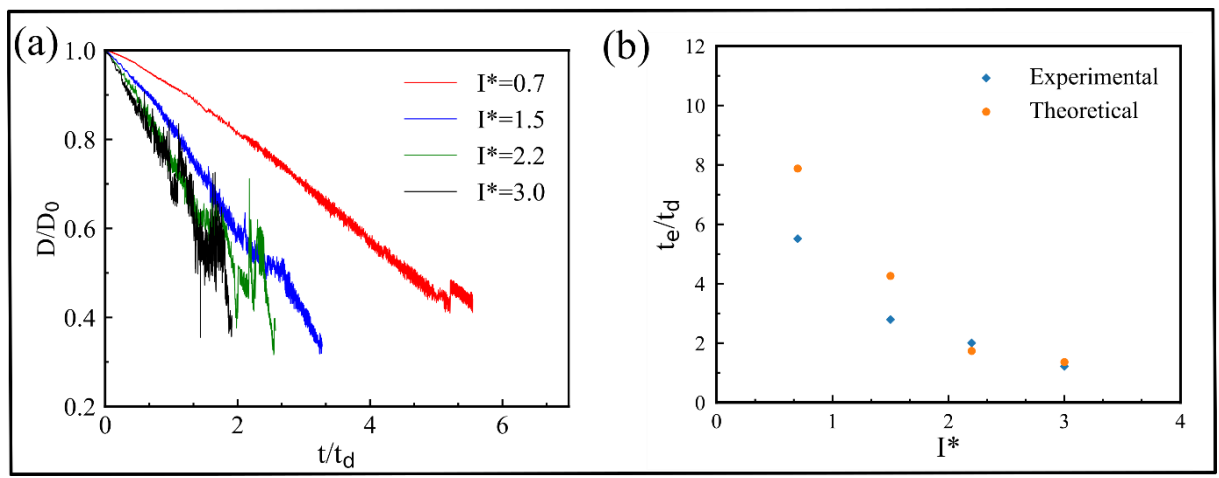

**Figure 7**  (a) Temporal evolution of droplet diameter at different laser irradiation intensities corresponding to $c/c^* = 14.1$ (b) Comparison of experimental and theoretical evaporation time scales at different irradiation intensities for $c/c^* = 14.1$.

Figure 7(a) shows the variation of normalized diameter ($D/D_0$) with normalized time ($t/t_d$) at different irradiation intensities for $c/c^* = 14.1$. As expected, diameter reduction is faster for high irradiation intensity and lower for lowest irradiation intensity. The time-varying amplitude of the evaporation curve represents bubble growth, whereas time-varying oscillations represent the droplet's rotational motion. The exact extraction of bubble growth scales, its variation with irradiation intensities, concentrations, and their theoretical comparisons will be elucidated in § 3.3. With regards to different concentrations at a particular irradiation intensity (see supplementary figure S4), the diameter reduction remains the same, implying evaporation time scales remain similar irrespective of polymer concentration. This enables a comparison of theoretical and experimental evaporation time scales for a particular concentration $c/c^* = 14.1$ at different irradiation intensities (see figure 7(b)). The theoretical evaporation time scales can be established from a diffusive law proposed by Sobac et al.(Sobac et al. 2019). The differential equation of the evolving drop radius can be written as

$$\frac{dD}{dt} = \frac{4}{D}\left(\frac{c_g d_{va}}{c_l}\right) \ln\left(\frac{1-X_i}{1-X_\infty}\right) \quad (3.5)$$

Using the initial condition $D(t = 0) = D_0$, the integration of the above equation gives

$$D(t)^2 = D_0^2 - 8d_{va}\left(\frac{c_g}{c_l}\right) \ln\left(\frac{1-X_\infty}{1-X_i}\right) t \quad (3.6)$$

where, $c_g$ represents gas molar concentration, $d_{va}$ represents the diffusion coefficient of vapor in air, $D$ represents the diameter of the drop, $X_i$ and $X_\infty$ represent the mole fraction of vapor in the gas phase at the interface and far field, respectively.

Using the slope of the above equation, an approximate theoretical evaporation time scale can be written as

$$t_{e,theoretical} \sim D_0^2/8d_{va}\left(\frac{c_g}{c_l}\right) \ln\left(\frac{1-X_\infty}{1-X_i}\right) \quad (3.7)$$

where $X_i$ can be written as

$$X_i = \exp\left(\frac{-L^*}{R_g}\left(\frac{1}{T_i} - \frac{1}{T_b}\right)\right) \quad (3.8)$$

where, $T_b$ is the boiling temperature of the liquid and $T_i$ is the interface temperature. $L^*$ is the molar latent heat of vaporization of the liquid and $R_g$ represents universal gas constant. Here $X_i$ is calculated using the saturation interface temperature obtained from infrared thermography. The experimental and theoretical evaporation time scales are in good agreement with each other, especially at higher irradiation intensities (see figure 7(b)). Note that the effect of acoustic streaming and skin layer formation will have an impact on evaporation time scales for $t_e > t_d$. However, for $t_e \sim t_d$, the evaporation time scales are governed by the diffusion time scales, perhaps indicating a close agreement of experimental and theoretical evaporation time scales at high irradiation intensities. Further, the characteristic frequency of evaporation is given by $f_e \sim 1/t_{e,theoretical}$ which is in $O(10^0)$ Hz. It closely matches the lower frequency band observed in figure 3(c), confirming that it represents evaporation.

## 3.3 Bubble growth (Phase B)

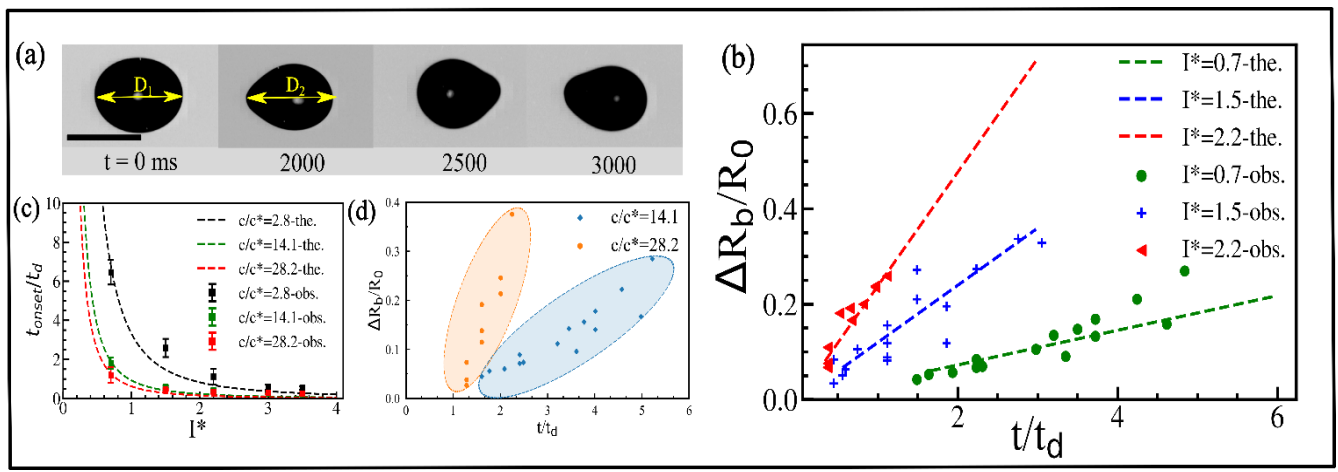

**Figure 8** (a) High-speed images of bubble growth for $c/c^* = 14.1$ and $I^* = 0.7$ (b) Effect of irradiation intensity on bubble growth scales and its theoretical comparison (c) Effect of irradiation intensity on the onset time of bubble growth at different irradiation intensities and its theoretical comparison (d) Effect of polymer concentration on the bubble growth at $I^* = 0.7$.

The energy interaction between the aqueous polymer droplet and the infrared laser is electromagnetic in origin. The droplet beam interaction could be understood based on the Mie-size parameter (Gannena et al. 2022) $2\pi R_0/\lambda_0$. For small value of Mie-size parameter, the source function (distribution of electromagnetic energy) distributed throughout the droplet volume is uniform corresponding to a uniform thermal energy evolution throughout the droplet. For large droplets, the Mie-size parameter increases resulting in the source function developing peaks (hot spots) and the thermal energy evolution shows inhomogeneities. Different hydrodynamic and thermodynamic processes are observed based on the internal thermal energy evolution inside the droplet. Bubble nucleation is an example of one such process. The aqueous polymer, being infrared opaque, interacts with the incident laser beam's electromagnetic field and absorbs most of the incident photons. The various energy interactions that occur during droplet beam interactions are radiative heating of the droplet, thermal diffusive processes throughout and at the interface evaporative and diffusive cooling of the droplet in general. However, the diffusive process occurring over the droplet length scale is typically slower than the radiative heating of the droplet. Therefore, at an intermediate and high intensity (evaporation time is shorter than the thermal diffusive time scale), the diffusive cooling process is too slow, and the temperature at the hotspots keeps increasing before nucleation occurs due to the formation of the skin layer at the droplet interface.

The average temperature of the droplet can be modelled using the energy equation with a volumetric source term. The volumetric source term represents the absorption process (Gannena et al. 2022). Droplets irradiated by lasers are typically in a metastable state. Therefore, the phase transition from liquid water to vapor does not occur at $100^0$ C. In case of rapid heating, the maximum temperature liquid water can achieve (i.e. the superheat limit) preserving its liquid state is close to $305^0$ C, beyond which spontaneous nucleation to vapor phase begins. The droplet hence can be in a liquid state beyond the boiling point at standard atmospheric pressure. As the temperature inside the droplet increases beyond the boiling point, and once the skin layer forms, bubble nucleation occurs much before the superheat temperature of liquid water is in a metastable state. The bubble growth in a viscoelastic medium is governed by the modified Rayleigh-Plesset equation (Gaudron et al. 2015).

$$R_b \ddot{R}_b + \frac{3}{2} \dot{R}_b^2 + \frac{4\nu_l \dot{R}_b}{R_b} + \frac{2\gamma}{\rho_l R_b} + \frac{E}{\rho_l} = \frac{p_B - p_\infty}{\rho_l} \tag{3.9}$$

where $R_b$ is the bubble radius, $\nu_l$ is the kinematic viscosity of the surrounding liquid around the bubble, $\gamma$ is the surface tension at the vapor-liquid interface, $\rho_l$ is the density of the liquid, and E is the elastic stress of the polymeric material. Usually, modelling the elastic stress for polymers mainly focuses on Maxwell-based models. However, systems that relax back to their original configuration are better modelled by Kelvin-Voigt polymeric models. Due to large deformations in polymeric systems, the infinitesimal strain assumption in most real-world scenarios is mainly invalid. This necessitates replacing the linear elasticity models with nonlinear finite strain elasticity models. Various nonlinear strain energy functions like neo-Hookean and Mooney-Rivlin approximations could be used. For the current experiments where the material relaxes back to its original state (PEO droplets, see supplementary Movie1 ) neo-Hookean models are much better. The elastic stress $E$ therefore, can be written as

$$E = \frac{\eta}{2} \left[ 5 - 4 \left( \frac{R_{b*}}{R_b} \right) - \left( \frac{R_{b*}}{R_b} \right)^4 \right] \tag{3.10}$$

Notice $E$ is directly proportional to the shear modulus $\eta$. $R_{b*}$ is the bubble size at the instant of nucleation and is typically of the order radius of gyration of polymer (It is in $O\ (nm)$).

In addition, for the bubble initiation phase, when the ratio $R_{b*}/R_b$ is close to unity, the correction terms are essential and elastic stress becomes time-varying and nonlinear. However, when the bubble has already grown to a size large enough such that the ratio $R_{b*}/R_b$ is much smaller than unity, the elastic stress does not dependent on the bubble radius and becomes a constant. Therefore, the elastic term becomes important only during the initial stage of the nucleation process. Further, owing to the very high ratio of the driving pressure $O\ (10^5$ Pa$)$ compared to the shear modulus $O\ (10^1$ Pa$)$, the elastic term in the Rayleigh-Plesset equation could be neglected (for low shear modulus). The right-hand side of the Rayleigh-Plesset equation is the driving pressure difference across the bubble and is of the order of the atmospheric pressure. $p_B$ is the pressure inside the bubble and $p_\infty$ is the pressure in the liquid phase (polymeric droplet). For an initial bubble size of $R_{b0} \gg R_{b*}$, the vapor pressure $p_{v0}$ inside the bubble is related to the bubble size $R_{b0}$ through the ideal gas equation (assuming the vapor inside the bubble to be an ideal gas) given by

$$p_{v0} R_{b0}^3 \propto \rho R_g T_\infty \tag{3.11}$$

where $\rho$ is the vapor density of water vapor, $T_\infty$ is the temperature of the liquid phase far away from the bubble, and $R_g$ is the gas constant. As the bubble expands, the bubble pressure varies with its size and gets coupled to the bubble temperature $T_B$ through the ideal gas law as given by

$$p_B R_b^3 \propto \rho R_g T_B \tag{3.12}$$

Dividing equation (3.12) by equation (3.11), we have

$$p_B = p_{v0} \left(\frac{T_B}{T_\infty}\right) \left(\frac{R_{b0}}{R_b}\right)^3 \tag{3.13}$$

The total pressure inside the bubble at any given time is given by

$$p_B(t) = p_v(T_B) + p_{v0} \left(\frac{T_B}{T_\infty}\right) \left(\frac{R_{b0}}{R_b}\right)^3 \tag{3.14}$$

Using equation (3.14) in equation (3.9), the right-hand side of the Rayleigh Plesset equation can be split into three terms as shown below

$$p_B(t) - p_\infty = I + II + III \tag{3.15}$$

where

$$I = \frac{p_v(T_\infty) - p_\infty}{\rho_l} \tag{3.16}$$

$$II = \frac{p_v(T_B) - p_v(T_\infty)}{\rho_l} \tag{3.17}$$

$$III = \frac{p_{v0}}{\rho_l} \left(\frac{T_B}{T_\infty}\right) \left(\frac{R_{b0}}{R_b}\right)^3 \tag{3.18}$$

Term I denotes the driving term in the far field region (far away from the bubble), term II represents the thermal term, and term III indicates the pressure inside the bubble.

If the liquid temperature is known, term I can be evaluated.

When the temperature difference between the bubble and the liquid temperature farther from the bubble $(T_B - T_\infty)$ is small, we can use Taylor series expansion to estimate term II. Keeping first-order quantities in $(T_B - T_\infty)$, term II becomes

$$\frac{p_v(T_B) - p_v(T_\infty)}{\rho_l} = B(T_B - T_\infty) \tag{3.19}$$

where the coefficient $B$ could be computed from the Claussius-Clapeyron equation.

The Claussius-Clapeyron equation is given by

$$\frac{dp}{dT} = \frac{L}{T \Delta v} \tag{3.20}$$

where $L$ is the latent heat of vaporization, $T$ is the temperature of phase change, $v$ is the specific volume. Integrating equation (3.20) and dividing by $\rho_l$ we have

$$\frac{p_v(T_B) - p_v(T_\infty)}{\rho_l} = \frac{\rho_v(T_\infty) L(T_\infty)}{\rho_l T_\infty}(T_B - T_\infty) \tag{3.21}$$

Comparing equation (3.21) with equation (3.19), we have

$$B = \frac{L}{\rho_l T_\infty} \rho_v(T_\infty) \tag{3.22}$$

During phase change at approximately atmospheric pressure, all the properties could be evaluated at the saturated temperature corresponding to atmospheric pressure (boiling point). The scale for $(T_B - T_\infty)$ is estimated using the thermal energy equations. In the current context, the liquid temperature is majorly dependent on radiative heating, as shown in our previous work at time scales smaller than the diffusive scales, which is valid for most bubble growth processes in the present context (Gannena et al. 2022). The average liquid temperature is therefore given by

$$T_l = T_0 + A \int_0^t G(R(t), \mu) dt \tag{3.23}$$

where

$$A = \frac{3\alpha I_0}{2\mu^3 \rho_l c_l} \tag{3.24}$$

and

$$G(R(t), \mu) = \frac{(\mu R + (\mu R - 1)e^{2\mu R} + 1)e^{-2\mu R}}{R^3} \tag{3.25}$$

Using equation (3.23), $(T_B - T_\infty)$ could be computed and can be approximated by the degree of superheat $(T_l - T_b) = \Delta T$ where $T_b$ is the boiling point of the liquid at atmospheric pressure. The energy balance at the bubble boundary $r = R_b$ is given by

$$4\pi R_b^2 L \rho_v(T_B) \dot{R}_b = 4\pi R_b^2 k_l \left(\frac{\partial T}{\partial r}\right)_{r=R_b} \tag{3.26}$$

where $k_l$ is the thermal conductivity of the liquid. The total energy equation is essentially nonlinear due to the bubble growth and coupling of $(T_B - T_\infty)$ and $R_b(t)$. Using the Plesset-Zwick approach (Plesset and Zwick 1954), the nonlinear terms are modelled based on the assumption that the thermal boundary layer is smaller than the bubble radius, i.e., $\delta_T \ll R_b(t)$. For the thin thermal boundary layer approximation, the unsteady thermal energy equation inside the liquid droplet can be solved with the Rayleigh-Plesset equation simultaneously in a coupled manner. On integrating the energy equation with respect to time and using the Rayleigh-Plesset equation,

The temperature difference $T_\infty - T_B(t)$ evaluated based on the Plesset Zwick criterion is given as

$$T_\infty - T_B(t) = \left(\frac{\alpha_l}{\pi}\right)^{1/2} \int_0^t \frac{R_b^2(x)\left(\frac{\partial T}{\partial r}\right)_{r=R_b(x)}}{(\int_x^t R_b^4(y) dy)^{1/2}} dx \tag{3.27}$$

where, $\alpha_l$ is the thermal diffusivity of the liquid. Further, $x$ and y represent dummy variables for the above-mentioned definite integral for time. We should notice that the capillary pressure term $(2\gamma/\rho_l R_b)$ does not appear in equation (3.27). This is due to the fact that the capillary term is approximately three orders of magnitude smaller than the right-hand side driving pressure in the Rayleigh-Plesset equation. For a bubble size $R_b \gtrsim R_{b0}$ that is bubble size of the order of 10 μm, the capillary pressure term is one order smaller than the driving term. Further, due to the inverse dependence of the capillary pressure term on bubble radius, the contribution of capillary pressure to the Rayleigh-Plesset equation decreases even further as the bubble expands. Using the boundary condition at the bubble interface from equation (3.26), equation (3.27) can be rewritten as

$$T_\infty - T_B(t) = \frac{L\rho_v}{\rho_l c_l \alpha_l^{1/2}} \left(\frac{1}{\pi}\right)^{1/2} \int_0^t \frac{R_b^2(x)\left(\frac{dR}{dt}\right)}{(\int_x^t R_b^4(y)dy)^{1/2}} dx \qquad (3.28)$$

In general, in most real-world scenarios, bubble growth can be approximated by a power law of the form

$$R_b = R^* t^n \qquad (3.29)$$

$$T_\infty - T_B(t) = \frac{L\rho_v}{\rho_l c_l \alpha_l^{1/2}} R^* t^{n-1/2} C(n) \qquad (3.30)$$

where

$$C(n) = n \left(\frac{4n+1}{\pi}\right)^{1/2} \int_0^1 \frac{z^{3n-1} dz}{(1-z^{4n+1})^{1/2}} \qquad (3.31)$$

where $0 < n < 1$. The Plesset-Zwick bubble growth curve is given by $n = 1/2$. Note for $n = 1/2$, $(T_\infty - T_B(t))$ is constant, indicating that both $T_\infty$ and $T_B$ change at the same rate. The approximate bubble growth length scale can also be obtained as a scaling consequence of equation (3.26) and employing the thin boundary layer assumption due to the Plesset-Zwick approximation. Equation (3.26) can be rewritten in terms of dominant scales as

$$4\pi R_b^2 L \rho_v(T_B) \dot{R}_b = 4\pi R_b^2 k_l \left(\frac{\partial T}{\partial r}\right)_{r=R_b} \sim 4\pi R_b^2 k_l \frac{\Delta T}{\delta_T} \qquad (3.32)$$

The thermal boundary layer scales as $\delta_T \sim \sqrt{\alpha_l t}$ in general. Using the time-varying thermal boundary length scale, the boundary condition at the wall can be written in terms of dominant scales as

$$L\rho_v(T_B) \frac{dR_b}{dt} \sim k_l \frac{\Delta T}{\sqrt{\alpha_l t}} \qquad (3.33)$$

Simplifying further, we have

$$dR_b \sim \frac{k_l \Delta T dt}{\rho_v(T_B)\sqrt{\alpha_l t}} \qquad (3.34)$$

Integrating equation (3.34), we have

$$\int dR_b \sim \int \frac{k_l \Delta T}{\rho_v(T_B)\sqrt{\alpha_l}} t^{-1/2} dt \qquad (3.35)$$

$$R_b - R_{b0} \sim \frac{2 k_l \Delta T}{\rho_v(T_B) L \sqrt{\alpha_l}} \frac{t^{1/2} \sqrt{\alpha_l}}{\sqrt{\alpha_l}} \qquad (3.36)$$

Dividing by $R_0$ and using $\alpha_l = k_l/\rho_l c_l$ we have

$$\frac{R_b - R_{b0}}{R_0} \sim \frac{2k_l \Delta T}{\rho_v(T_B) L \alpha_l} \left(\frac{\alpha_l t}{R_0^2}\right)^{1/2} \tag{3.37}$$

Simplifying further and recognizing the diffusion time scale $t_d$ we have

$$\frac{\Delta R_b}{R_0} \sim \frac{\rho_l c_l \Delta T}{\rho_v L} \left(\frac{t}{t_d}\right)^{1/2} \tag{3.38}$$

The bubble radius scale can also be obtained using equation (3.29) and equation (3.30)

$$R_b \sim \frac{1}{C(1/2)} \frac{\rho_l c_l \Delta T}{\rho_v L} (\alpha_l t)^{1/2} \tag{3.39}$$

where $C(1/2)$ could be computed from equation (3.31).

After the initial fast bubble growth phase near the nucleation point, the bubble growth rate relatively slows down and grows as $R_b \sim (t)^{1/2}$ (refer to eq 3.39). As the bubble grows to a size comparable to the droplet radius, a noticeable bump can be observed on the illumination face of the droplet. The bubble always appears on the illumination face of the droplet because the temperature of the droplet is higher on the illumination face compared to the shadow face. The temperature decreases exponentially as the laser beam traverses from the illumination to the shadow face due to Beer-Lamberts law (Gannena et al. 2022). The time it takes to observe a detectable bump in the droplet is known as the onset time ($t_{onset}$) and the corresponding droplet length scale is known as onset radius ($R_{onset}$) or diameter ($D_{onset}$). The approximate length scale of the bubble could be computed using a unique feature of the experimental configuration. Generally, the droplet in a levitated field rotates about its center of mass axis due to the conservation of angular momentum. Figure 8(a) shows the droplet and the bump caused by the bubble growing inside during various phases of its rotation. Once the bubble nucleates and grows, we observe a bump on the illumination face of the droplet ($t = 2000\ ms$ here) and measure its horizontal length scale $D_2$. As the droplet rotates, we see the view as it appeared at $t = 0\ ms$. Thus, an approximate bubble length scale is estimated as $D_2 - D_1$. Note that the bubble remains stationary once it nucleates due to the low Reynolds number (Re) flow inside the droplet. The approximate liquid Reynolds number can be determined as follows. From the mass conservation boundary condition at the bubble liquid interface, the approximate liquid velocity ($V_l$) can be written as $V_l \sim V_b(\rho_v/\rho_l)$, where $V_b \sim \frac{dR_b}{dt}$ and $\rho_v$ represents bubble velocity and vapor density, respectively. The Reynolds number ($Re \sim \rho_l V_l D_{onset}/\mu_l$) comes out to be in $O(10^{-7})$ implying negligible flow conditions inside the droplet. The experimental bubble length scale is plotted in figure 8(b). The theoretical scale of bubble growth according to equation (3.38) is plotted, and it agrees well with the experimental data within the experimental uncertainty range for various irradiation intensities. Higher values of $I^*$ corresponds to a higher degree of superheat for various $I^*$. The degree of superheat enters the bubble growth equation in the coefficient of $(t)^{1/2}$. The onset times ($t_{onset}$) can also be related to the irradiation intensity $I^*$ through the bubble growth equation (3.38) and through the degree of superheat $\Delta T$. The degree of superheating is related to $I^*$. This can be understood from the equation of the liquid temperature (3.23).

$$\Delta T \propto A \propto I_0 \tag{3.40}$$

From equation (3.38), we have,

$$R_b \propto \Delta T t^{1/2} \tag{3.41}$$

Using equation (3.40) in equation (3.41) and realizing that at the onset time $R_b \propto R_{onset}$ we have

$$t_{onset}^{1/2} \propto \frac{R_{onset}}{I_0} \tag{3.42}$$

Simplifying further, we have

$$t_{onset} \propto \frac{R_{onset}^2}{I_0^2} \tag{3.43}$$

The irradiation intensity can be expressed in its non-dimensional form as

$$I^* = \frac{1.5 D_0 I_0}{h_{lv} \alpha_l \rho_l} \tag{3.44}$$

where $D_0$ is the initial diameter of the droplet, $h_{lv}$ represents latent heat of vaporization.

$$t_{onset} \propto (R_{onset}/I^*)^2 \tag{3.45}$$

The theoretical $t_{onset}$ curves obtained for various non-dimensional irradiation intensities and concentrations pass closely through the experimentally observed $t_{onset}$ values (see figure 8(c)). The constant of proportionality for equation (3.45) is obtained by calibrating the onset radius and onset time scale appearing in equation (3.45) at the lowest value of $I^*$. Furthermore, at a specific irradiation intensity, the bubble growth scales are higher for larger $c/c^*$ compared to a lower $c/c^*$ (see figure 8(d)). The critical gelation concentration $\emptyset_g$ to form a skin layer reaches much earlier in the evaporation time scale, thus leading to earlier nucleation of bubble and subsequently larger bubble growth scales for high $c/c^*$ compared to lower $c/c^*$ at a particular irradiation intensity.

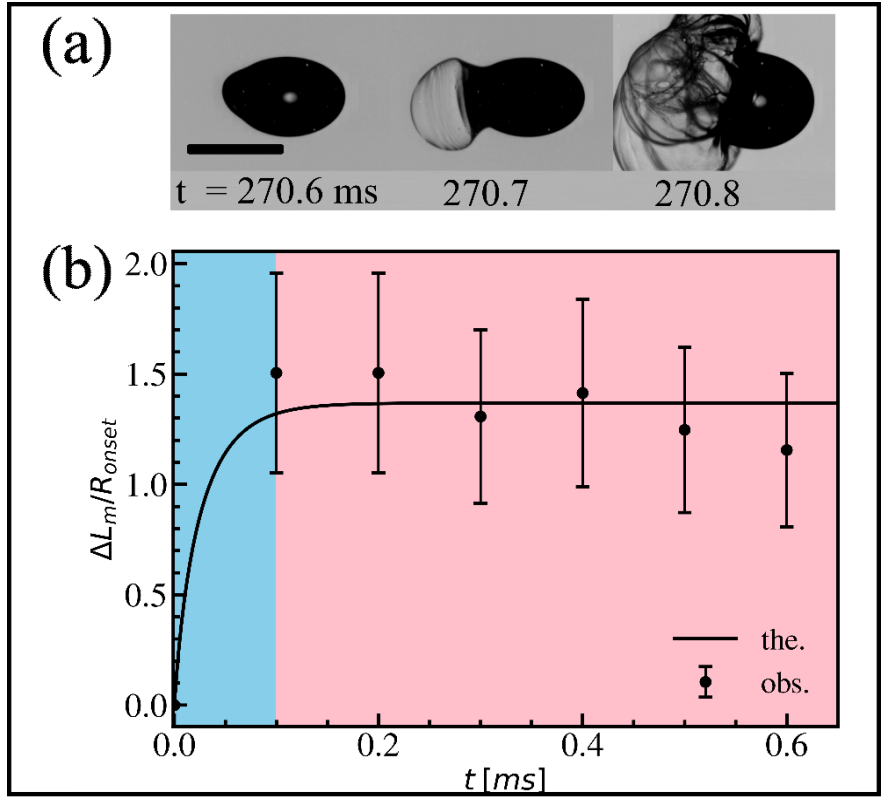

**Figure 9** (a) High-speed images of membrane growth for $c/c^* = 28.2$ and $I^* = 3.5$ (b) Membrane growth dynamics at $c/c^* = 28.2$, $I^* = 3.5$ and its theoretical comparison. The scale bar represents 1 mm.

We model the membrane growth dynamics (see figure 9(a)) using a spring, mass, and damper model with external forcing from the pressure difference across the membrane interface coupled to the maximum pressure inside the bubble through the Clausius-Clapeyron equation. The governing dynamical law for the viscoelastic membrane is given as (Gannena et al. 2022)

$$\ddot{L}_m + 2\zeta \omega_n \dot{L}_m + \omega_n^2 L_m = \frac{C}{R_m} \tag{3.46}$$

where

$$C = \frac{3 m_{in} R_g T_B}{16 \pi \rho_m h_0 R_{onset}^2} \tag{3.47}$$

$$\omega_n \sim \sqrt{\frac{2E}{\rho_m R_{onset}^2}} \tag{3.48}$$

The parameters $C$ and $\omega_n$ are computed using the onset radius ($R_{onset}$), membrane thickness ($h_0$), and the temperature inside the bubble ($T_B$). The bubble temperature is approximately the same as the liquid temperature ($T_l$) owing to the Plesset Zwick criteria given by equation (3.30) for $n = 1/2$, and hence the liquid temperature computed by using equations (3.23, 3.24, and 3.25) is used. For most parametric values of the irradiation intensities ($I^* = 0.7 - 2.2$) quasi steady bubble growth is observed for all values of polymeric concentration considered in this study. However, for extreme irradiation intensities ($I^* = 3, 3.5$), the bubble pressure rises rapidly due to the rapid temperature rise. Owing to the bubble's rapid expansion rate and pressure, the polymeric membrane expansion becomes highly transient in contrast to the quasi-steady nature of bubble growth for low irradiation intensities. Using $E \sim O(10^9)$ Pa (Jee et al. 2013) and $\zeta \sim O(10^2)$ in equation (3.46) and solving for the growth length scale, we get a growth time scales of 0.1 ms which is of the same order of magnitude as the experimental growth rate of the polymeric membrane. Note that $E > \Delta P$ (where $\Delta P$ represents pressure difference across the membrane which is in $O(10^5)$ Pa) leads to membrane expansion and collapse without rupture (see supplementary Movie1 ). Figure 9(b) shows the comparison of the theoretical membrane growth scale according to equation (3.46) with the experimental membrane growth length scale ($L_m$) which is in reasonable agreement within the experimental uncertainty. Note that due to the complicated shape of the membrane expansion, the experimental length scales are extracted manually.

### 3.4 Shape oscillations and precipitate formation (Phases C and D)

The shape oscillations are majorly driven by the presence of a bubble inside the droplet and, to an extent, by the rotational motion of the droplet.

The bubble growth close to the illuminating face of the droplet causes the evaporating polymer droplet to become asymmetric. This enabled us to quantify the rotational motion of the evaporating polymer droplet in a levitation field. The time-varying frequencies of the rotational motion of the droplet are obtained by performing power spectrum operation on the diameter within short time intervals. Once we know the temporal variation of diameter and frequency of droplet rotation, the relation between them in a levitated experimental configuration can be obtained as follows. From the principle of angular momentum conservation of the droplet (assuming negligible losses due to air resistance at the air-droplet interface)

$$I\omega = k \tag{3.49}$$

where $I = mD^2/10$ represents the moment of inertia of the droplet, m is the mass of the droplet, D is the diameter of the droplet, and $\omega = 2\pi f$ represents the frequency of rotation of the droplet.

$$\frac{mD^2}{10} 2\pi f = k \tag{3.50}$$

$$\frac{\rho_l \pi^2 f D^5}{30} = k \tag{3.51}$$

$$\frac{D^5(t)}{D^5(t_{ref})} = \frac{f(t_{ref})}{f(t)} \tag{3.52}$$

$$f(t) = f(t_{ref}) \left(\frac{D(t_{ref})}{D(t)}\right)^5 \tag{3.53}$$

$$f(t) \propto \frac{1}{D^5(t)} \tag{3.54}$$

Here $f(t_{ref})$ and $D(t_{ref})$ represents frequency and diameter at the reference initial time scale in the evaporation process of the droplet. The relation between rotation frequency and droplet diameter is implicitly independent of concentration and rotation. Hence the experimental and theoretical comparison of diameter variation with rotational frequency of droplet is carried out for $c/c^* = 14.1$ and $I^* = 1.5$. The experimental and theoretical values show reasonable agreement within the experimental uncertainty (see figure 10). Further, the theoretical frequency scale closely matches the higher frequency band in the power spectrum of diameter regression (see figure 3(c)).

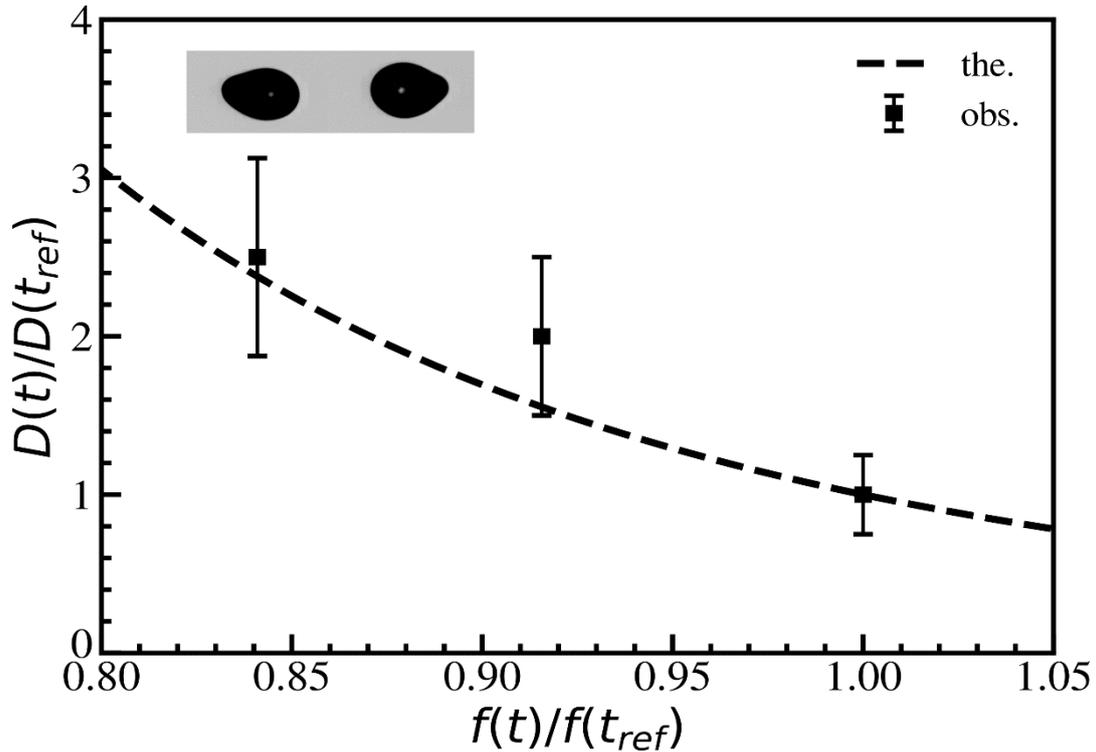

**Figure 10** Experimental and theoretical comparison of variation of rotational frequency of the droplet with diameter for $c/c^* = 14.1$ and $I^* = 1.5$.

Further, we observe vigorous droplet shape oscillations primarily due to the presence of bubble. The droplet shape oscillations are quantified by estimating the change in the droplet position (x-coordinate and y-coordinate) with time. The shape oscillations over a wide range of irradiation intensities and concentrations are characterized through a non-dimensional parameter referred as bubble growth index ($\alpha$), which is defined as $\alpha = (D_{b,max}/D_{onset})^3$. Where $D_{b,max}$ represents the maximum expansion diameter of the polymer droplet due to the bubble growth and $D_{onset}$ represents the diameter of the polymer droplet at the onset of nucleation. It is observed that the major shape oscillations are characterized by $\alpha >= 1$, which are predominant at high concentrations and high irradiation intensities ($c/c^* = 14.1, 28.2$ and $I^* = 3, 3.5$). Similarly, mild shape oscillations are characterized by $0.5 <= \alpha < 1$. These small-scale oscillations are dominant at high concentrations and low irradiation intensities ($c/c^* = 14.1, 28.2$ and $I^* = 0.7 - 2.2$). Finally, $\alpha < 0.5$ signifies minor shape oscillations observed at low concentrations and all irradiation intensities($c/c^* <$ 10 and $I^* = 0.7 - 3.5$). Figure 11(a) shows a typical major shape oscillation event. The oscillations start when the bubble/membrane expands and collapses into the polymer droplet. The major shape oscillations are characterized by volumetric shape distortions, stretching and reorientation of the polymer droplet (see inset figure 11(a)). Figure 11(a) depicts the typical centroid trajectory of the polymer droplet when the membrane (bubble) collapses back into the parent droplet. The normalized centroid coordinates of the polymer droplet are defined as $X^* = (X - X_{onset})/X_{onset}$ and $Y^* = (Y - Y_{onset})/Y_{onset}$ where $X_{onset}$ and $Y_{onset}$ represents centroid coordinates of the polymer droplet when membrane (bubble) collapse occurs. As evident from figure 11(a), once the membrane (bubble) collapses into the parent droplet, the droplet oscillates vigorously along both the horizontal and vertical directions. The maximum non-dimensional displacement in the centroid's positive x-coordinate and y-coordinate position is 2.4 and 0.4, respectively, while the non-dimensional displacement in the negative coordinates is −2 and −0.6, respectively. In contrast, centroid displacement for mild to minor shape oscillations is comparatively smaller than for major shape oscillations. For mild shape oscillations, the maximum non-dimensional displacement in the positive x-coordinate and y-coordinate position of the centroid is 0.1 and 0.15, respectively, while the non-dimensional displacement in the negative coordinates is – 1.5 and −0.3, respectively (see supplementary figure S5). Whereas for very mild shape oscillations, the maximum non-dimensional displacement in the centroid's positive x-coordinate and y-coordinate position is 1.6 and 0.15, respectively, while the non-dimensional displacement in the negative coordinates is – 1.3 and −0.15, respectively (see figure 11(b)). In the case of $\alpha > 1$, nucleation of multiple bubbles and their subsequent coalescence leads to a complex 'Dumbell' shape. This leads to droplets experiencing vigorous motion in both horizontal and vertical directions. In the regime $\alpha < 1$, the expanding bubble remains stationary close to the illumination face of the droplet for most of the droplet evaporation lifetime, resulting in only minor shape oscillations. Power spectrum density operation on X-center of mass (X-CM) and Y-center of mass (Y-CM) revealed that the dominant frequencies of droplet oscillations are in $O(10^1 - 10^2)$ Hz. The frequencies remain independent of concentration and irradiation intensity, implying that these oscillations are driven by levitation system parameters contrary to the amplitude of the oscillations (trajectories), which are guided by the bubble growth and evaporation dynamics.

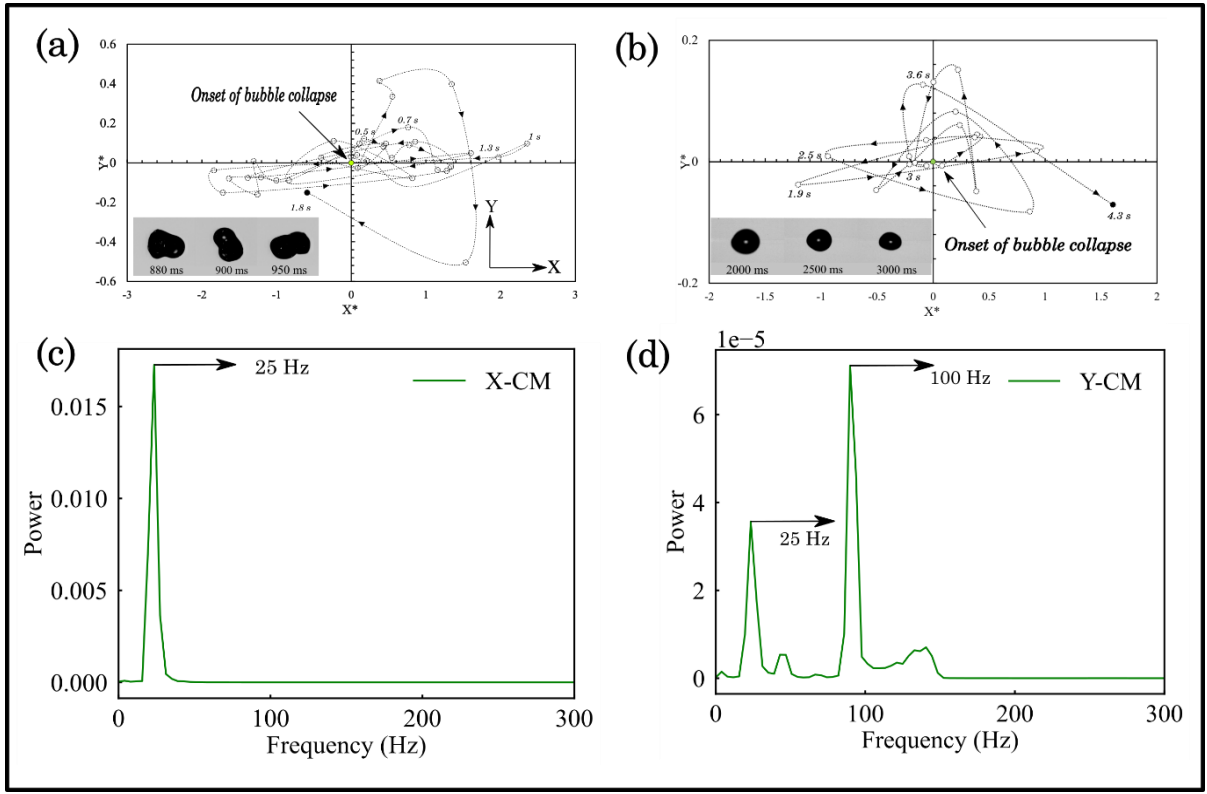

**Figure 11** (a) Typical trajectory of the centroid of levitated polymer droplet for $\alpha > 1$. (b) Typical trajectory of the centroid of levitated polymer droplet for $0 < \alpha < 0.5$. (c) Power spectrum density of X–center of mass (X-CM) of evaporating polymer droplet at $I^* = 1.5$ and $c/c^* = 14.1$ (d) Power spectrum density of Y–center of mass (Y-CM) of evaporating polymer droplet at $I^* = 1.5$ and $c/c^* = 14.1$. The frequencies of centroid motion remain the same irrespective of concentration and irradiation intensity.

| Bubble growth index | Onset of bubble /membrane growth | Shape oscillations | Interaction of multiple bubbles | Final strcture | Maximum occurence probability |
|---|---|---|---|---|---|
| $\alpha>1$ | | | | | $c/c^*>10$ and at high irradiation intensities |
| $0.5<\alpha<1$ | | | | | $c/c^*>10$ and at low irradiation intensities |
| $\alpha<0.5$ | | | | | $c/c^*<10$ and at all irradiation intensities |

**Figure 12** Summary depicting the effect of nucleated bubbles on underlying polymer droplet dynamics and final precipitates in different regimes.

When the concentration of PEO is large ($c/c^* > 10$) and at high irradiation intensities, a significant amount of nucleation sites are formed, resulting in large bubble formation. The expanding bubble eventually stretches the skin layer and expands as a viscoelastic membrane (Gannena et al. 2022). Due to significantly higher elasticity, the membrane stretches and collapses into the parent droplet. These events are observed intermittently throughout the evaporation phase in the event $\alpha > 1$. For $\alpha < 1$, due to smaller bubble growth, the unique membrane expansion and collapse is not observed. The bubble size substantially affects the polymer droplet's final morphology. At high concentrations and irradiation intensities, where the bubble diameter is significantly large ($\alpha > 1$), the final structure of the polymer droplet resembles a shell structure (figure 12). For $\alpha < 0.5$, where the size of the bubble is significantly smaller owing to fewer nucleation sites, the final structure is a smooth solid precipitate. The bubble nucleation in multi-component, polymeric droplets with the low viscoelastic modulus (PAM) results in different modes (ligament mediated, catastrophic, micro-explosion) of atomization of droplets which has not been observed in polymeric droplets with the high viscoelastic modulus (PEO) even at significantly high concentrations and irradiation intensities. The significantly higher strength of the skin layer hampers the droplet atomization process.

## 4. Conclusions

A comprehensive experimental investigation is performed to understand the bubble dynamics and droplet shape oscillations in acoustically levitated polymer droplets under external radiative heating. The conclusions derived from the present work are as follows:

1. High viscoelastic modulus droplets experience evaporation and droplet shape oscillations without nucleation-induced atomization, in contrast to the occurrence of previously reported distinct modes of atomization in low viscoelastic modulus polymer droplets. This is attributed to the increased entanglement density at the skin layer, which directly correlates with the higher skin layer strength. The steeper increase in entanglement density ($N_e$) for High viscoelastic modulus (PEO) droplets leads to the nucleation of vapor bubbles in dilute, semi-dilute unentangled regimes contrary to the occurrence of bubble nucleation in low viscoelastic modulus (PAM) droplets in the semi-dilute entangled regime.

2. Depending on laser irradiation intensity and polymer concentration, four temporal phases are observed: droplet evaporation (phase A), vapor bubble nucleation followed by bubble/membrane growth (phase B), shape oscillations and precipitate formation (phases C and D). The time scale for the droplet evaporation phase is in $O(10^0 - 10^1)$ s. The theoretical time scale obtained from a diffusive evaporative law predicts the experimental evaporation time scale. This subsequently predicts a low-frequency band observed in the power spectrum of diameter regression.

3. The scaling analysis shows that the quasi-steady bubble length scale $R_b$ varies temporally as $(t)^{1/2}$ for low irradiation intensities. The membrane growth dynamics are modelled at high irradiation intensities using a spring mass damper system. There is good agreement between the theoretical ($O(10^{-4})s$) and experimental ($O(10^{-4})s$) growth time scales. Further, it is observed that the onset time ($t_{onset}$) of bubble growth varies with irradiation intensity as $t_{onset} \propto (R_{onset}/I^*)^2$.

4. From the principle of conservation of angular momentum, it is shown that the frequency of rotation varies with droplet diameter as $f \propto 1/D^5$. In addition, the theoretical rotation frequency agrees well with the high-frequency band observed in the power spectrum of diameter regression.

5. Finally, a bubble growth index $\alpha$ is defined, which characterizes the final precipitates formed after evaporation. $\alpha > 1$ and $0.5 < \alpha < 1$ is characterized by shell-like precipitates, whereas for $\alpha < 0.5$, solid precipitates are observed. These findings are contrary to the different atomization modes observed for low viscoelastic modulus polymer droplets in similar concentration regimes, further deciphering the role played by the skin layer in the dynamics of evaporating polymeric droplets.

## Declaration of Interests

The authors declare no conflict of interest.